\documentclass[10pt,twocolumn,letterpaper]{article}

\usepackage{cvpr}
\usepackage{times}
\usepackage{epsfig}
\usepackage{graphicx}
\usepackage{amsmath}
\usepackage{amssymb}

\usepackage[breaklinks=true,bookmarks=false]{hyperref}

\cvprfinalcopy 

\def\cvprPaperID{****} 

\begin{document}

\title{To believe or not to believe:\\ Validating explanation fidelity for dynamic malware analysis}

\author{Li Chen\\
li.chen@intel.com \\
Intel Labs\\
\and
Carter Yagemann \\
yagemann@gatech.edu\\ 
Georgia Institute of Technology
\and 
Evan Downing \\
edowning3@gatech.edu\\
Georgia Institute of Technology
}

\maketitle

\begin{abstract}

Converting malware into images followed by vision-based deep learning algorithms has shown superior threat detection efficacy compared with classical machine learning algorithms. When malware are visualized as images, visual-based interpretation schemes can also be applied to extract insights of why individual samples are classified as malicious. In this work, via two case studies of dynamic malware classification, we extend the local interpretable model-agnostic explanation algorithm to explain image-based dynamic malware classification and examine its interpretation fidelity. For both case studies, we first train deep learning models via transfer learning on malware images, demonstrate high classification effectiveness, apply an explanation method on the images, and correlate the results back to the samples to validate whether the algorithmic insights are consistent with security domain expertise. In our first case study, the interpretation framework identifies indirect calls that uniquely characterize the underlying exploit behavior of a malware family. In our second case study, the interpretation framework extracts insightful information such as cryptography-related APIs when applied on images created from API existence, but generate ambiguous interpretation on images created from API sequences and frequencies. Our findings indicate that current image-based interpretation techniques are promising for explaining vision-based malware classification. We continue to develop image-based interpretation schemes specifically for security applications.
\end{abstract}

\section{Introduction}

Malware is malicious software created for harming users, computers, and networks. Viruses, trojan horses, worms, spyware, and ransomware are examples of malware. In malware detection, static analysis without executing the application is a quick method to detect malicious patterns in an application. To avoid static detection most malware contain obfuscated code. Dynamic analysis, on the other hand, executes the code and records the malware runtime behavior. Even though dynamic analysis is slower than static analysis, it offers better resiliency and efficacy against malware code obfuscation. 

Machine learning has become increasingly popular and important for malware detection because it can generalize to detect new malware families. The manual effort of feature engineering can be costly, especially on unstructured data formats. As the volume of data continues to grow at increasing speed, scalable algorithms for malware detection are in high demand.
Computer vision has provided a unique perspective for performing malware classification. First, it enables natural visualization on malware as a whole entity. Second, deep learning has demonstrated state-of-the-art performance for image classification. When malware is represented as images, transfer learning can leverage the superior performance from vision to classify malware with accelerated training speed and maintained classification efficacy. Last but not least, it has superior performance compared with classical machine learning algorithms~\cite{nataraj2011malware, makandar2015malware, yue2017imbalanced, chen2018henet, chen2018deep}.

For static malware analysis, a binary can be directly mapped to pixel values between 0 and 255~\cite{nataraj2011malware, makandar2015malware, yue2017imbalanced, chen2018henet, chen2018deep}. By visually inspecting binaries plotted as grey-scale images, we can observe the textural and structural similarities or dissimilarities on the \textit{static} features of malware.
By contrast, there are fewer vision-based~\textit{dynamic} malware classification techniques. \cite{chen2018henet} proposed a hierarchical ensemble neural network scheme on dynamic telemetries collected from Intel\textsuperscript{\textregistered} Processor Trace, where the control flow packets are converted into time series of images and demonstrated the superior performance compared with other popular dynamic malware classifiers.  

For security applications, besides classification efficacy, model explanation is equally important for security researchers and practitioners to deploy the model in the wild. Sensible interpretation from the model on why a sample is predicted as malicious or benign can generate valuable insights to triage malware families, identify new malware signatures, understand the evolution of polymorphic malware, and enhance the practitioner's trust in the model. When malware is represented as images, interpretation schemes for natural images~\cite{ribeiro2016should} can be extended to explain malware classification. 

Unlike natural images, where interpretation fidelity can be assessed via human eyes, interpretation fidelity on malware images remains to be validated through security domain expertise. 
In this paper, via two case studies for dynamic malware classification, we investigate the effectiveness of local-interpretable model-agnostic explanation (LIME) framework~\cite{ribeiro2016should} specifically for image-based dynamic malware analysis. Our first case study examines dynamic malware images generated from predictions on sequences of instructions. The interpretation framework identifies indirect calls that uniquely characterize the underlying exploit behavior of a malware family. In our second case study, we consider three types of malware images generated from API existence, API sequence, and API frequency features. The interpretation framework provides insightful information such as crypto-related APIs when applied on images created from API existence, but generates ambiguous information on images created from API sequences and frequencies. Our findings indicate that current image-based interpretation techniques are promising for vision-based malware classification. We plan to develop image-based interpretation schemes specifically for malware images in security applications.


Our contributions are summarized as follows:
\begin{itemize}
    \item To the best of the authors' knowledge, we are the first to validate the interpretation fidelity of a model-agnostic interpretation framework, using security domain expertise, on dynamic image-based malware classification.
    \item We use deep transfer learning on dynamic malware images generated from instruction sequence predictions, API existence, API sequence, and API frequency features and demonstrate that dynamic malware image analysis is highly effective.
    \item Our case studies present a valuable combination of machine learning and domain expertise to fully understand the effectiveness of malware classification algorithms.
    \item We advocate that interpretation is another important dimension to evaluate malware classifiers. Vision-based interpretability highlights the advantage of approaching the malware problem from a computer vision direction so that interpretation becomes concrete as to indicate the actual locations of potential malicious signals. 
\end{itemize}

\section{Background and Related Work}


In the interpretation frameworks for image classification, the explanation method provides interpretation by identifying the most contributing pixel regions to the prediction result~\cite{fong2017interpretable, gehr2018ai2, lundberg2017unified, ribeiro2016should}. 
While there is an abundance of vision and natural language based interpretation frameworks, few exist specifically for security applications. In~\cite{guo2018lemna}, the authors proposed non-linear approximation on the local decision boundary to explain malware detection algorithms for security applications. The method is primarily for multi-layer perception (MLP) and recurrent neural networks (RNN) on non-image based data representations for malware classification. \cite{chen2018deep} employed interpretation frameworks such as the local-interpretable model-agnostic explanation (LIME)~\cite{ribeiro2016should} for natural images on static malware images.

When we represent dynamic malware as images, natural image explanation schemes can be applied. In two case studies here, we extend LIME to image-based dynamic malware classification and thoroughly examine the interpretation fidelity using security domain knowledge. 
\section{Case Studies}

\subsection{Case Study I}
Our first case study is concerned with detecting anomalies in dynamic program control-flow traces. 
The task is to examine whether PDF files opened by Adobe Acrobat Reader are malicious or not in a Windows\textsuperscript{\textregistered} system. The source of our control-flow data is Intel\textsuperscript{\textregistered} Processor Trace (Intel\textsuperscript{\textregistered} PT), which is a hardware feature present in modern Intel\textsuperscript{\textregistered} processors~\cite{chen2018henet}. Intel\textsuperscript{\textregistered} PT produces a large volume of data within a short time period. For example, tracing Acrobat Reader for one minute yields over \textit{2 million} indirect control-flow transfers including returns and indirect calls and jumps. It becomes a daunting task for human analysts to examine such a high volume of transfers for signs of exploitation. Hence it is desired to employ the automated interpretation framework to extract an explanation. 

We collect 1,249 benign and 1,314 malicious traces from the \textit{pdfka} malware family, where each trace is collected from the targeted program opening a PDF document. Each trace is then disassembled, yielding a linear sequence of the executed basic blocks. A basic block is defined as a sequence of linear instructions ending with a branch, which can be a return, call, jump, conditional branch and so on. Each basic block is assigned a universally unique integer defined as BBID. A fixed length sliding window is moved over the sequence of BBIDs, and a subsequent long short term memory (LSTM) neural network is tasked with learning and predicting the next BBID for any sub-sequence ending with an indirect control-flow transfer. The intuition behind predicting only indirect transfers is that these are the only places where control-flow hijacking can occur during a program execution. The LSTM model is trained using only normal traces of the target program and then its performance is monitored over unlabeled traces. If an anomaly occurs in the trace, this will cause the model's performance to drop below a defined threshold and the trace will be labeled anomalous.

The dynamic malware images are generated from the prediction of the LSTM model on the BBIDs, where the white pixels are correct predictions and black pixels are incorrect predictions. 
On these malware images, we apply deep transfer learning using the pre-trained VGG model~\cite{simonyan2014very} on ImageNet, freeze the top layers and add an additional two fully connected layers, each with dropout, to retrain on the dynamic malware images. The training and test split is $0.8:0.2$. We set the number of epochs to be 50 with early stopping criterion if the validation loss does not decrease after 10 epochs. We use the model checkpointed at the 32-th epoch. The classification accuracy on the test set is 100\%. This result demonstrates the effectiveness of vision-based deep transfer learning approach for dynamic malware detection and thus it makes sense to examine what interpretation can be generated using the decision boundaries from this model. 

\begin{figure}
\centering
\includegraphics[width=0.5\textwidth]{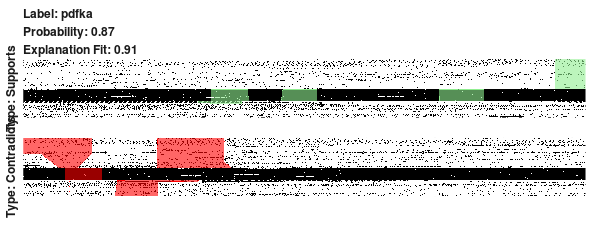}
\caption{\label{fig:pt-malicious} Interpretation for a \textit{pdfka} sample.
White pixels are correct predictions by the underlying model and black incorrect. Green denotes
strong support whereas red is strong contradiction. The green regions shown highlight a suspicious control-flow loop.}
\end{figure}

Next we apply the interpretation framework on the generated images to examine the fidelity of explanation. Figure~\ref{fig:pt-malicious} shows the interpretation of one of the \textit{pdfka} family traces. Our model marks several spots within a large streak of incorrect predictions as strongly supporting that this trace belongs to the \textit{pdfka} set. From here we can reverse these chunks of BBIDs to get back to the executed instructions. Upon manual inspection, we discover that at this point in the trace, the program makes one particular indirect call several hundred times in a row. By looking at the relative virtual address, we determine that this activity is happening inside the part of \textit{AcroRd32.dll} that parses TIFF images. The most well-known vulnerability in this part of the program is CVE-2010-0188, which matches several online reports about this family's behavior~\cite{PdfkaReport}. We also manually confirm that this pattern does in fact appear in all the \textit{pdfka} traces and none of the benign traces. To further verify, we also create and trace several benign documents containing benign TIFF images and confirm none of them produce the anomalous pattern. Although Acrobat Reader is closed source, making indisputable verification difficult, we believe our manual analysis strongly supports that our interpretation model successfully identified the subsection of the \textit{pdfka} traces that uniquely characterizes the underlying exploit this family relies on. This case study demonstrates the usefulness of the interpretation method on dynamic malware images.

\subsection{Case Study II}

In our second case study, we evaluate three models designed to classify Windows\textsuperscript{R} Portable Executable (PE) files as either malicious or benign. All three models use dynamic features produced by malware and benign software during execution. 

Our malware dataset is comprised of 13,394 Windows\textsuperscript{R} PE samples. These samples were collected from the Georgia Tech Research Institute (GTRI) using their internal malware collection and analysis platform APIARY~\cite{APIARY}. Using AVClass~\cite{sebastian_avclass:_2016}, our malware dataset is made up of 247 families (demonstrating the diversity of our samples).

Our benign dataset is made up of 5,772 samples and was collected by crawling CNET~\cite{CNET}. Specifically, our samples are a mix of Windows\textsuperscript{R} PE and Windows\textsuperscript{R} Installer (MSI) files under 22 different categories (according to CNET) ranging from Audio to Education to Business-related software.

We ran all of our samples for 2 minutes using a modified version of Cuckoo~\cite{Cuckoo} version 1.2 in Windows 7 32-bit KVM virtual machines with network and random-access memory (RAM) hardware extensions. We used KVM and hardware extensions to introduce as few artifacts indicative of a malware analysis environment as possible. Malware authors have been known to check for system and network-related artifacts (e.g., registry key values and network timing) which they can use to evade analysis (e.g., by performing innocuous activities or terminating early)~\cite{kolbitsch_power_2011,kirat_barecloud:_2014,dinaburg_ether:_2008,royal_entrapment:_2012}. To improve the quality of our malware dataset, we only include samples which ran for the full 2 minutes without terminating early. We also executed the malware samples with 3 days of them being collected by the organization to improve the chances that the malware would perform malicious activities. Finally, to improve the quality of our benign dataset, we only include samples which did not have more than 2 antivirus companies label them as malicious via VirusTotal~\cite{VirusTotal}. We also use Cuckoo to automatically interact with the benign software (via fake mouse-clicks on its GUI) to cause it to reveal a variety of behaviors (namely the installation process).

On each set of the dynamic malware images generated from API existence, sequence, and frequency, we again apply transfer learning via VGG network pre-trained on ImageNet, where we freeze the top layers and add customized two additional fully connected layers and a softmax layer to produce the classification result. The training and validation split is 0.8:0.2. We set the number of epochs to be 50 with early stopping criterion if the validation loss does not decrease after 10 epochs. The classification results on three models are the same within statistically significance with an accuracy at 95\%. Then we apply the interpretation scheme on each of the three datasets and the decision boundaries generated by the corresponding classifiers. 


The interpretation framework on API call frequencies and existence generate similar insights, where we find that one of our malicious sample interpretations focused on cryptography-related API calls and HTTP-related calls, both of which are common ways for malware to communicate with their command-and-control (C\&C servers). We note that these extracted insights are not exclusive to malware, since legitimate Internet browser applications perform similar activities. In fact, this lends insight into the shortcomings of our benign dataset and what types of benign software we may be missing to improve the reliability of our classifier.

\begin{figure}
\centering
\includegraphics[width=0.5\textwidth]{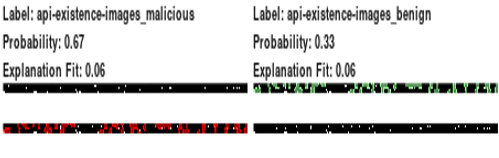}
\caption{\label{fig:existence:malicious} Interpretation for a dynamic malicious image created from API existence. The large light green areas in the top left image denote support regions for the malicious class. The supported regions that contribute to malicious prediction focused on cryptography-related API calls and HTTP-related calls.}
\end{figure}

\begin{figure}
\centering
\includegraphics[width=0.5\textwidth]{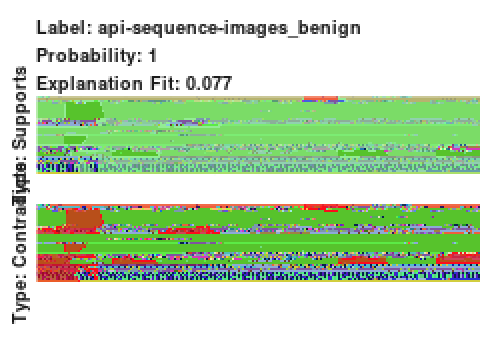}
\caption{\label{fig:seq:benign} Interpretation for a benign sample. Each color represents a unique Windows API call during execution. The large light green areas in the top image denote support for the benign class. The dark red areas in the bottom image contradict the support for the benign class.}
\end{figure}

\begin{figure}
\centering
\includegraphics[width=0.5\textwidth]{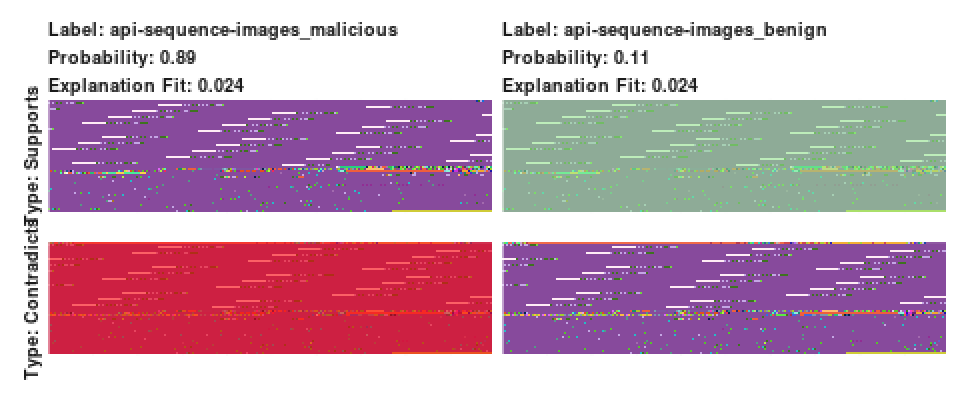}
\caption{\label{fig:seq:malicious} Interpretation on a malicious image generated from API sequences, where each pixel represents a unique Windows API call during execution. Even though the vision-based  classification scheme correctly predicts this sample as malicious with high confidence, the interpretation method that approximates the boundaries provides ambiguous explanation.}
\end{figure}

The API sequence call has been shown to be a weak feature in past work~\cite{wagner_mimicry_2002}. Using the interpretation framework, we examine whether the API sequence is a weak feature without relying on domain expertise. Although training this model resulted in a validation accuracy of 94\%, our interpretation results are not intuitive. While the model was confident in classifying one the benign samples (Fig~\ref{fig:seq:benign}), the interpretation on its boundary approximations is ambiguous at interpreting why this was the case. There are large sections highlighted as contributing to the classification of the sample. The most heavily weighted areas make frequent calls to \textit{FindNextFileW} and \textit{GetProcAddress} (among others), but this isn't indicative of benign \emph{or} malicious behavior. When looking at malicious samples, the results are even more ambiguous. It seems the model memorized at least one of the samples entirely as seen in Fig~\ref{fig:seq:malicious}.

\section{Conclusion}
In this paper, we demonstrate the effectiveness of using computer-vision based techniques for dynamic malware classification and employing vision-based interpretation frameworks to explain why the deep learning models make such predictions. Our discoveries on the two case studies indicate the promising advantages of applying vision-based interpretation frameworks to explain image-based dynamic malware classifiers. Security practitioners, based on the algorithmic interpretation findings, can check the code and verify whether the ML-identified locations contain a signatures unique to certain malware families. We plan to continue studying and proposing interpretation schemes specifically for image-based malware classification frameworks.


{\small
\bibliographystyle{ieee}
\bibliography{egbib}
}

\end{document}